\documentstyle[12pt]{article}
\def\be{\begin{equation}}
\def\ee{\end{equation}}
\begin{document}

\hfill McGill/92-31

\hfill IFT-P.025/92

\hfill July 1992

\hfill revised Nov.~1992
\vskip 1.5cm
{\Large \center Lepton Masses in an $SU(3)_L \otimes U(1)_N$ Gauge Model \\}
\vskip 2cm
{\center {\bf R. Foot$^{a}$, O. F. Hern\'andez$^{b}$, F. Pisano$^{c}$,
and V. Pleitez$^{c}$}\\
\vskip .8cm
{$a$: \ } Physics Department, Southampton University\\
Highfield, Southampton, SO9 5NH, U.K.\\
\vskip .5cm
{$b$: \ } Physics Department, McGill University\\
3600 University St., Montr\'eal, Qu\'ebec, Canada H3A 2T8.\\
email: oscarh@physics.mcgill.ca
\vskip .5cm
{$c$: \ } Instituto de F\'\i sica Te\'orica\\
Universidade Estadual Paulista, Rua Pamplona, 145\\
01405-900--S\~ao Paulo, SP, Brasil.\\}
\vskip 1cm
{\center Abstract\\}
\vskip .5cm

The $SU(3)_c \otimes SU(3)_L \otimes U(1)_N$ model of Pisano and Pleitez
extends the Standard Model in a particularly nice way, so
that for example the anomalies cancel only when the number of
generations is divisible by three. The original version of the model
has some problems accounting for the lepton masses. We resolve this
problem by modifying the details of the symmetry breaking sector in
the model.

\newpage

In references \cite{0},\cite{1}, two of us proposed a model
based on the gauge symmetry:
\be
SU(3)_c \otimes SU(3)_L \otimes U(1)_N.
\label{group}\ee
In those original papers spontaneous symmetry breaking
and fermion mass generation are assumed to arise from
the vacuum expectation values (VEV) of three scalar multiplets,
$\chi, \rho$ and $\eta$ which are each triplets under $SU(3)_L$.
Here we would like to point out that these scalar multiplets do not
give satisfactory masses to the leptons and we
resolve the problem by modifying the details of the symmetry breaking
sector. We then verify that this modification does not change the
model's attractive feature or its compatibility with experiment.

We first give a brief review of the model.
The three lepton generations transform under the gauge symmetry,
Eq.~(\ref{group}), as
\be
f^a = \left(
\begin{array}{c}
\nu_L{}^a\\
e_L{}^a\\
e^c_R{}^a
\end{array}\right)_L \sim (1, 3, 0) \  ,
\label{leptons}\ee
where $a=1,2,3$ is the generation index.

Two of the three quark generations transform identically and one
generation, it does not matter which, transforms in a different
representation of $SU(3)_L \otimes U(1)_N$.
Thus we give the quarks the following representation under
Eq.(\ref{group}):
$$
Q_{1L} = \left(
\begin{array}{c}
u_1\\
d_1\\
J_1
\end{array}\right)_L \sim (3, 3, 2/3),
$$
$$
u_{1R} \sim (3, 1, 2/3), \
d_{1R} \sim (3, 1, -1/3),  \  J_{1R} \sim (3, 1, 5/3)
$$
$$Q_{2L} = \left(
\begin{array}{c}
d_{2}\\
u_{2}\\
J_{2}
\end{array}\right)_L \sim (3, 3^*, -1/3),
$$
$$u_{2R} \sim (3, 1, 2/3), \  d_{2R} \sim (3, 1, -1/3), \
J_{2R} \sim (3, 1, -4/3)
$$
$$
Q_{3L} = \left(\begin{array}{c}
d_3\\
u_3\\
J_{3}
\end{array}\right)_L \sim (3, 3^*, -1/3),
$$
\be
u_{3R} \sim (3, 1, 2/3),\
d_{3R} \sim (3, 1, -1/3), \  J_{3R} \sim (3, 1, -4/3)
\label{quarks}\ee
One can easily check that all gauge anomalies cancel
in this theory. However, note that each generation is anomalous.
In fact this type of construction is only anomaly free when
the number of generations is divisible by 3. Thus 3 generations is
singled out as the simplest non-trivial anomaly free $SU(3)_L \otimes
U(1)_N$ model.

We introduce the Higgs
\be
\chi \sim (1, 3, -1),
\label{chihiggs}\ee
which couples via the Yukawa Lagrangian:
\be
{\cal L}_{yuk}^{\chi} = \lambda_1 \bar Q_{1L} J_{1R} \chi
+ \lambda_{ij} \bar Q_{iL} J_{jR} \chi^* + H.c.
\label{chiyuk}\ee
where $i,j = 2,3$. If $\chi$ gets the VEV
\be
\langle \chi \rangle =
\left(\begin{array}{c}
0\\
0\\
w
\end{array}\right)
\label{chivev}\ee
the exotic charged $5/3$ and $-4/3$ quarks ($J_{1,2,3}$) gain mass
and the gauge symmetry is broken:
\be\begin{array}{c}
SU(3)_c \otimes SU(3)_L \otimes U(1)_N \\
\downarrow \langle \chi \rangle\\
SU(3)_c \otimes SU(2)_L \otimes U(1)_Y
\end{array}
\label{chibreak}\ee
Even though the model has charged $5/3$ and $-4/3$ quarks there
will be no fractional charged color singlet bound states,
and hence no absolutely stable fractionally charged particles in the model.

The usual standard model $U(1)_Y$ hypercharge is given by
\be
Y = 2N - \sqrt{3} \lambda_8 \ .
\label{hyper}\ee
Here $\lambda_8$ is the Gell-Mann matrix diag[1,1,-2]$/\sqrt{3}$.
The model reduces to the standard
model as an effective theory at a intermediate scale.

In the original papers \cite{0},\cite{1}, electroweak symmetry breaking and
fermion masses was assumed to be due to the scalar bosons
\be\rho \sim (1, 3, 1), \ \eta \sim (1, 3, 0) \label{rhoeta} \ee
These scalar bosons couple to the fermions through the
Yukawa Lagrangians:
\be
{\cal L}_{yuk}^{\rho} = \lambda_{1a} \bar Q_{1L} d_{aR} \rho
                + \lambda_{ia} \bar Q_{iL} u_{aR} \rho^*
                + H.c.
\label{rhoyuk}\ee
\be
{\cal L}_{yuk}^{\eta} = G_{ab}\bar f_{aL} (f_{bL})^c \eta^* +
\lambda'_{1a} \bar Q_{1L} u_{aR} \eta +
\lambda'_{ia} \bar Q_{iL} d_{aR} \eta^* + H.c.
\label{etayuk}\ee
where $a,b=1,2,3$ and $i=2,3$.
When the $\rho$ gets the VEV:
\be
\langle \rho \rangle = \left(
\begin{array}{c}
0\\
u\\
0
\end{array}\right)
\label{rhovev}\ee
two up and one down type quark gain mass.  The down quark that gets its mass
from the $\rho$ is not the isospin partner of the two other up quarks.

If $\eta$ gets the VEV:
\be
\langle \eta \rangle = \left(
\begin {array}{c}
v\\
0\\
0
\end{array}\right),
\label{etavev}\ee
then the remaining quarks get mass.
However not all of the leptons get mass. This is because the first
term in Eq.(\ref{etayuk}) is only non-zero when $G_{ab}$ is
antisymmetric in the generation indices ($a,b$). To see this note
that the Lorentz contraction is antisymmetric, and the fields are
Grassman (so that this gives a antisymmetric factor when they are
interchanged) and the $SU(3)_L$ contraction is antisymmetric.
Explicitly writing the $SU(3)_L$ indices the leptonic term in
eq.(\ref{etayuk}) we have
\be
G_{ab}\bar f_{iaL} (f_{jbL})^c \eta_k^* \epsilon^{ijk}
\label{explicit}\ee
We have three antisymmetric factors
hence only the antisymmetric part of the coupling
constants $G_{kl}$ gives a non-vanishing contribution
and the mass matrix for the leptons is antisymmetric.
A $3\times 3$ antisymmetric mass matrix has eigenvalues
$0, -M, M$, so that one of the leptons does not gain mass
and the other two are degenerate, at least at tree level.

The simplest way to remedy this situation is to modify the symmetry breaking
sector of the model.
If the leptons are to get their masses at tree level within the
usual Higgs mechanism, then
we need a Higgs multiplet which couples to
$\bar f_L (f_L)^c$.
Since
\be\bar f_L (f_L)^c \sim (1, 3 + 6^*, 0),
\label{ffc} \ee
then the only scalars which can couple to $\bar f_L (f_L)^c$
must transform as a  $(1, 3^*, 0)$ or $(1, 6, 0)$ (or the complex
conjugate there-of). The simplest choice was the $(1, 3^*, 0)$ option
which failed due to the fact that the $3\times 3 \times 3 \ SU(3)$
invariant is antisymmetric. However the $6$ is a symmetric
product of $3 \times 3$, and it can couple to $\bar f_L (f_L)^c$,
so it seems that a Higgs multiplet
$S \sim (1, 6, 0) $ can give the leptons their masses.

The VEV of $S$ must have the form
\be\langle S \rangle =
\left(\begin{array}{ccc}
0&0&0\\
0&0&{v'\over\sqrt2}\\
0&{v'\over\sqrt2}&0
\end{array}\right) \label{svev} \ee
Note that when the VEV has this form, it gives the leptons their
masses and together with $\langle \rho \rangle,\langle \eta \rangle$
breaks the electroweak gauge symmetry:
\be\begin{array}{c}
SU(3)_c \otimes SU(2)_L \otimes U(1)_Y\\
\downarrow \langle \rho \rangle, \langle \eta \rangle ,\langle S \rangle\\
SU(3)_c \otimes U(1)_Q
\end{array} \label{breaking} \ee

It is now no longer obvious that the Higgs potential can be chosen in such a
way that the all the Higgs fields get their desired VEVs.  We must show two
things.  First that there exists a range of values for the parameters in the
Higgs potential such that the VEVs given by
Eqs.~(\ref{chivev}),~(\ref{rhovev}),~(\ref{etavev}),~(\ref{svev}) give a local
minimum. And secondly that the number of Goldstone bosons that arise from the
symmetry breaking in the scalar field sector of the theory is exactly equal to
eight.  This will ensure that there are no pseudo-Goldstone bosons arising from
the breaking of a global symmetry in the scalar sector which is larger
than the $SU(3)\otimes U(1)$ gauge symmetry.

The Higgs potential has the form:
\begin{eqnarray}
V(\eta,\rho,\chi,S)&=& \lambda_1 [ \eta^\dagger \eta - v^2]^2
+\lambda_2 [\rho^\dagger \rho -u^2 ]^2
+\lambda_3 [ \chi^\dagger \chi - w^2 ]^2
\nonumber \\ & &\mbox{}
+ \lambda_4 [  Tr(S^\dagger S)   - v'^2 ]^2
+ \lambda_5 [ 2 Tr(S^\dagger S S^\dagger S) - (Tr[S^\dagger S])^2 ]
\nonumber \\ & &\mbox{}
+\lambda_6 [ \eta^\dagger \eta - v^2 + \rho^\dagger \rho -u^2 ]^2
+\lambda_7 [ \eta^\dagger \eta - v^2 + \chi^\dagger \chi -w^2 ]^2
\nonumber \\ & &\mbox{}
+\lambda_8 [ \chi^\dagger \chi - w^2 + \rho^\dagger \rho -u^2 ]^2
+ \lambda_9 [ \eta^\dagger \eta - v^2 + Tr(S^\dagger S) - v'^2]^2
\nonumber \\ & &\mbox{}
+ \lambda_{10} [ \rho^\dagger \rho - u^2 + Tr(S^\dagger S) - v'^2]^2
+ \lambda_{11} [ \chi^\dagger \chi - w^2 + Tr(S^\dagger S) - v'^2]^2
\nonumber \\ & &\mbox{}
+ \lambda_{12} [ \rho^\dagger \eta ] [\eta^\dagger \rho]
+ \lambda_{13} [ \chi^\dagger \eta ] [ \eta^\dagger \chi ]
+ \lambda_{14} [ \rho^\dagger \chi ] [ \chi^\dagger \rho ]
\nonumber \\ & &\mbox{}
+ f_1 \epsilon_{i,j,k} \eta_i \rho_j \chi_k
+ f_2 \rho^T S^\dagger \chi + H.c.
\label{potential}
\end{eqnarray}

A detailed analysis
shows that for all $\lambda$'s$>0$,
there exist values of $f_1$ and $f_2$ such that
the potential is minimized by the desired VEVs and such that there are no
pseudo-Goldstone bosons.  The above potential leads exactly to 8 Goldstone
bosons which are eaten by the gauge bosons which acquire a mass.
However even without a rigorous analysis one expects
such a result to be true for the following reasons.
If the terms $\lambda_5$,
$f_1$ and $f_2$ are zero, the above potential is positive definite
and zero when
\be
\langle \eta \rangle =
\left(\begin {array}{c}
v\\
0\\
0
\end {array}
\right), \
\langle \rho \rangle = \left(\begin {array}{c}
0\\
u\\
0
\end {array}
\right), \
\langle \chi \rangle = \left( \begin {array}{c}
0\\
0\\
w
\end {array}
\right),\
\langle S \rangle =
\left(\begin{array}{ccc}
0&0&0\\
0&0&{v'\over\sqrt2}\\
0&{v'\over\sqrt2}&0
\end{array}\right)
\label {e99}
\ee

Hence the above VEVs are a minimum of the potential. Note that the
$\lambda_{12}, \lambda_{13}$ and $\lambda_{14}$ terms in the Higgs potential
are very important for the alignment of the vacuum, as they imply that the
three vectors $\langle \eta \rangle, \langle \rho \rangle,$ and $\langle \chi
\rangle$ are orthogonal (in the complex 3 dimensional mathematical space).

Now allow $\lambda_5$ to be non-zero and positive.  This term is positive in
most of the parameter space of the matrix $S$.  For it to be negative we must
have large values of $Tr[S^\dagger S]$ which in turn would imply large values
for terms like $\lambda_4, \lambda_{10}, \lambda_{11}$.   Unless we allow fine
tuning of the potential so that $\lambda_4, \lambda_{10}, \lambda_{11}$ are
very small, the desired $S$ VEV minimizes the potential.  However any $S$ VEV
such that $Tr[S^\dagger S]=v'^2$ will minimize the potential and make it
zero.
Now consider non-zero
$f_1$, $f_2$.  These trilinear terms ensure that the largest continuum symmetry
of the scalar potential is $SU(3)\otimes U(1)$. In addition to this the $f_2$
term is linear in $S$ and this term induces a VEV for $S$ proportional to the
VEV of $\langle \rho \chi^T \rangle$ which has the desired form (given in
Eq.(\ref {svev})).

The potential in Eq.~(\ref{potential}) is
the most general $SU(3)\otimes U(1)$ gauge invariant,
renormalizable Higgs potential for the three triplets and the sextet,
which also respects the following discrete symmetry
\be
 \rho \to i\rho, \ \chi \to i\chi, \ \eta \to -\eta, \  S \to -S
\label{discrete}
\ee
If the fermions transform as
\be
f_L \to i f_L, \ Q_{1L} \to -Q_{1L}, \  Q_{iL} \to -iQ_{iL},
 \ u_{aR} \to u_{aR}, \ d_{aR} \to id_{aR}
\label{disfer}
\ee
the entire Lagrangian is kept invariant.
This symmetry is important since it prevents the trilinear terms
\be
\eta^TS^{\dagger}\eta \ \  {\rm and} \ \
  \epsilon^{ijk}\epsilon^{lmn}S_{il}S_{jm}S_{kn}
\label{numasstri}
\ee
from appearing in the Higgs potential.  These terms make analysis of
the Higgs potential more complicated and lead to nonzero
Majorana neutrino masses.
To see that without the discrete symmetry Majorana neutrino masses would occur
define
\be
S =
\left(\begin{array}{ccc}
\sigma^0_1&h^-_2&h^+_1\\
h^-_2&H^{--}_1&{\sigma^0_2}\\
h^+_1&{\sigma^0_2}&H^{++}_2
\end{array}\right) \ . \label{svev2} \ee
Note that $S$ couples with
leptons via the Yukawa Lagrangian
\begin{eqnarray}
2{\cal L}_{lS}&\!\!=\!\!&-\sum_lG_l[(\bar\nu^c_{lL}\nu_{lL}\sigma_1^0+\bar
l^c_Ll_LH_2^{++}+ \bar l_Rl^c_LH_1^{--})
+(\bar\nu^c_{lR}l_L +\bar l^c_R\nu_L)h_1^+\nonumber \\ &&\mbox{}
+(\bar\nu^c_{lR}l^c_L+\bar l_R\nu_{lL})h_2^-
+(\bar l^c_Rl^c_L+\bar l_Rl_L)\sigma_2^0]+H.c.
\label{lepyuk}
\end{eqnarray}
The neutrino gets a Majorana mass if $\langle
\sigma_1^0\rangle\not=0$ and it is this VEV which the symmetry
(\ref{discrete}), (\ref{disfer}) keeps equal to zero by preventing the terms
in (\ref{numasstri}).
If we did not impose the symmetry we could always fine tune
$\langle \sigma_1^0\rangle$ to zero, but this is a more unattractive option.

Note that the VEVs of $\rho,\eta$ and $S$ break the gauge
symmetry while preserving the tree-level mass relation
$M_W^2 = M_Z^2 \cos^2\theta_W$. (Here we define $\sin\theta_W\equiv
e/g$, the ratio of the electric charge coupling to the $SU(2)_L$
coupling after $\chi$ acquires a VEV.) One way to see this is to note that
under the intermediate scale gauge group $SU(2)_L \otimes U(1)_Y$,
the VEVs of $\rho$ and $S$ transform as members of a $Y = 1$,
$SU(2)_L$ doublet, and the custodial $SU(2)_C$ symmetry is not
broken.
One can also see this explicitly by working out the vector
bosons masses. The mass matrix for the neutral gauge boson is the following
in the $(W^3,W^8,B_N)$ basis:
\be
M^2=\frac{1}{4}g^2
\left(\begin{array}{lll}
a+b+a'           & \frac{1}{\sqrt3}(a-b+a')     & -2tb\\
\frac{1}{\sqrt3}(a-b+a')&\frac{1}{3}(a+b+4c+a') &\frac{2}{\sqrt3}t(b+2c)\\
-2tb               &\frac{2}{\sqrt3}t(b+2c)  &4t^2(b+c)
\end{array}\right)
\label{m}
\ee
with the notation
\[a=2v^2,\,b=2u^2,\,c=2w^2,\,a'=2v'^2,\,t=g_N/g_{SU(3)_L} \ . \]
We can verify that ${\rm det}M^2=0$.

The eigenvalues of the matrix in Eq.~(\ref{m}) are $0,M^2_Z$ and $M^2_{Z'}$
respectively. In the approximation that $c>>a,b,a'$, $M^2_Z$ and
$M^2_{Z'}$ become
\be
M^2_Z=\frac{g^2}{4}(a+b+a')\frac{1+4t^2}{1+3t^2}
\label{apzmass}
\ee
\be
M^2_{Z'}=\frac{g^2}{3}(1+3t^2)c,
\label{apzpmass}
\ee
Hence we can see that $M^2_{Z'}$ is very massive since it depends
only on $w$. In Ref.~\cite{1} a lower bound of 40 TeV have
been obtained by considering the contribution to the $K^0-\bar K^0$
mass difference of the heavy $Z'^0$.

There is a charged gauge bosons with mass given by
\be
M^2_W=\frac{1}{4}g^2(a+b+a')
\label{apwmass}
\ee
Then, as in Ref.~\cite{1}
\be
\frac{M^2_Z}{M^2_W}=\frac{1+4t^2}{1+3t^2}
\label{apcos}
\ee
In order to get consistency with experimental data Eq.~(\ref{apcos})
must be numerically equal to $1/\cos^2_W$, where $\theta_W$ is the
weak mixing angle in the standard electroweak model.
This definition of the weak mixing angle is
consistent with the previous definition (i.e. $\sin\theta_W= e/g$)
at tree level since we can always define
\[t^2=\frac{s^2_W}{1-4s^2_W} \ \  . \]
This shows that at tree level the $\rho$ parameter is equal to one as in the
standard model.

We have reviewed the model proposed in \cite{0},\cite{1} and have shown how to
modify it to yield a realistic lepton mass spectrum at tree level.  We have
proven that a Higgs potential allowing such a change exists and we have
verified that such a modification does not lead to any problems such as a
tree level $\rho$-parameter different from one.

The model has many unique features. In particular it is
only anomaly free if the number of
generations is a multiple of three.
Models of this type deserve our attention and study.

{\bf \center Acknowledgements \\}

OFH was supported by the National Science and Engineering Research Council of
Canada, and les Fonds FCAR du Qu\'ebec. FP and VP would like to thank the
Con\-se\-lho Na\-cio\-nal de De\-sen\-vol\-vi\-men\-to Cien\-t\'\i \-fi\-co e
Tec\-no\-l\'o\-gi\-co (CNPq) for full (FP) and partial (VP) financial
support.

\bibliographystyle{unsrt}

\begin{thebibliography}{99}

\bibitem{0}
F. Pisano and V. Pleitez, ``Neutrinoless double beta decay
and doubly charged gauge bosons'', IFT-P.017/91, July 1991.

\bibitem{1}
F. Pisano and V. Pleitez, ``$SU(3)\otimes U(1)$ model for electroweak
interactions'', Phys. Rev. {\bf D46}, 410(1992).

\end{thebibliography}

\end{document}